\documentclass[10pt]{IEEEtran}
\usepackage[final]{graphicx}
\usepackage{subfigure}
\usepackage{cite,amsmath,amsthm,amssymb,amsfonts,calrsfs}
\usepackage{srcltx}

\newcommand{\comment}[1]{}
\newcommand{\field}[1]{\mathbb{#1}} 

\newtheorem{theorem}{Theorem}

\makeatletter
\def\ps@headings{%
\def\@oddhead{\mbox{}\scriptsize\rightmark \hfil \thepage}%
\def\@evenhead{\scriptsize\thepage \hfil \leftmark\mbox{}}%
\def\@oddfoot{}%
\def\@evenfoot{}}
\makeatother 
\def\figref#1{Fig.~\ref{#1}}
\def\thref#1{Theorem~\ref{#1}}

\def\secref#1{Sec.~\ref{#1}}

\begin{document}

\title{Diversity-Multiplexing Tradeoff of Cooperative
    Communication with Linear Network Coded Relays}
\author{\normalsize
\begin{tabular}{ccc}
Hakan Topakkaya   \hspace{1mm} and \hspace{1mm} Zhengdao Wang\\
\small Dept. of Elec. and Comp. Eng.,
 \small Iowa State University,
 \small Ames, IA 50011-3060 \small \\
Emails: \{hakan, zhengdao\}@iastate.edu
\end{tabular}
}

\maketitle

\begin{abstract}

Network coding and cooperative communication have received
considerable attention from the research community recently in
order to mitigate the adverse effects of fading in wireless
transmissions and at the same time to achieve high throughput and
better spectral efficiency. In this work, we analyze a network
coding scheme for a cooperative communication setup with multiple
sources and destinations. The proposed protocol achieves the full
diversity order at the expense of a slightly reduced multiplexing
rate compared to existing schemes in the literature. We show that
our scheme outperforms conventional cooperation in terms of the
diversity-multiplexing tradeoff.
\end{abstract}

\section{Introduction} \label{intro}

Channel fading is one significant cause of performance degradation
in wireless networks. In order to combat fading, diversity
techniques that operate in time, frequency or space are commonly
employed. The basic idea is to send the signals that carry same
information through different paths, allowing the receiver to
obtain multiple independently faded replicas of the data symbols.
Cooperative diversity tries to exploit spatial diversity using a
collection of distributed antennas belonging to different
terminals, hence creating a virtual array rather than using
physical arrays.

In \cite{acly00} Ahlswede et al. introduced \emph{network coding}
to achieve the max-flow value for single-source multicast which
could be impossible by simply routing the data. Since then,
network coding has been recognized as a useful technique in
increasing the throughput of a wired/wireless network. The basic
idea of network coding is that an intermediate node does not
simply route the information but instead combines several input
packets from its neighbors with its own packets and then forwards
it to the next hop. However, since network coding is devised at
the network layer, error-free communication from the physical and
medium-access layer is usually assumed, which is a simplifying
assumption for wireless communications.

Efforts have also been made to apply network coding to the
physical layer, e.g. in \cite{zhll06}. Towards that goal,
cooperative schemes have been proposed that make use of network
coding in a cooperative communication setup, and studies have been
conducted to determine whether network coding provides any
advantages over existing cooperative communication techniques
\cite{yulb07,pzzy08}.

In \cite{pzzy08}, a network-coded cooperation (NCC) was proposed
and its performance was quantified using the
diversity-multiplexing tradeoff analysis which was originally
proposed for multiple antenna systems in \cite{zhts03}. NCC was
shown to outperform conventional cooperation (CC) schemes which
includes space-time coded protocols \cite{lawo03} and selection
relaying \cite{bkrl06}. NCC requires less bandwidth, and yields
similar or reduced system outage probability while achieving the
same diversity order. However, these results are based on an
optimistic assumption that any destination node should receive the
packets that are not intended for it without any error so that the
intended packet can be recovered from the xor'ed packet sent by
the relay. When this assumption is removed the scheme can no
longer achieve the full diversity order of $M+1$, where $M$ is the
number of cooperating relays, but only a reduced diversity order
of 2.

In this paper, we propose a network coded cooperation schemes for
$N$ source-destination pairs assisted with $M$ relays. The
proposed scheme allows the relays to apply network coding on the
data it has received from its neighbors using the coefficients
from the parity-check matrix of a $(N+M,M,N+1)$ MDS code. A closed
form expression for the outage probability is derived. We also
obtain the diversity-multiplexing tradeoff performance of the
proposed scheme. Specifically, it achieves the full-diversity
order $M+1$ at the expense of a slightly reduced multiplexing
rate. We show that our scheme outperforms NCC and CC in terms of
the probability of outage.


The rest of the paper is organized as follows. Section
\ref{sec.sys} discusses the system model, description of the
proposed scheme and the design of the network code. In Section
\ref{sec.dmt}, we present the main result and the proof. In
Section \ref{sec.compare} the performance of the proposed scheme
is compared in terms of DMT and average outage probability.
Section \ref{sec.conclusion} contains the conclusions.
\begin{figure}
\centering
\includegraphics[width=.25\textwidth]{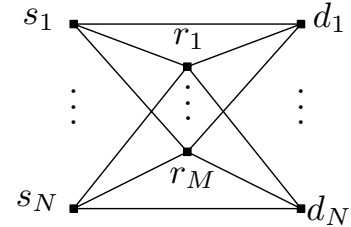}
\caption{System model: $N$ source-destination pairs and $M$
relays} \label{network-cap.pdf}
\end{figure}
\section{System Model} \label{sec.sys}

\subsection{General System Description}

\begin{figure}
\centering
\includegraphics[width=.45\textwidth]{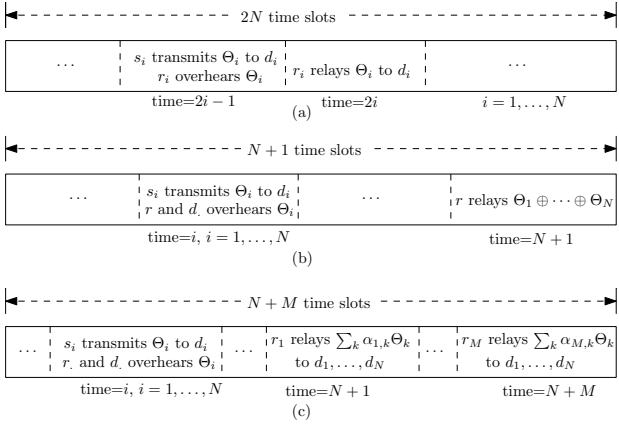}
\caption{time-division allocation for the different schemes
compared: (a) CC (b) NCC (c) DNCC} \label{fig.ta}
\end{figure}

The network studied in the paper is composed of $N$
source-destination pairs denoted as $(s_1,d_1), \ldots,
(s_N,d_N)$, and $M$ relays denoted as $r_1, \ldots, r_M$ in a
single-cell where all the nodes can hear the transmissions of each
other as shown in \figref{network-cap.pdf}. We consider two
different transmission scenarios. In the first scenario, each
source node $s_i$ is trying to transmit the data packet $\theta_i$
to all the destinations $d_i, i=1, \ldots, N$ which is known as
the \emph{multicast} scenario. In the second scenario, each source
node $s_i$ is trying to transmit the data packet $\theta_i$ to
only destination $d_i$ and we will refer to this scenario as the
\emph{unicast} scenario. We assume that each packet is composed of
$L$ bits: $\theta_i = [\theta_{i, 1}, \theta_{i,2}, \ldots,
\theta_{i, L}]$. As in \cite{kome03}, these length-$L$ blocks of
bits that are transmitted on each link will be treated as elements
of a finite field $\mathbb{F}_q$, $q=2^L$. All the nodes are
assumed to be equipped with half-duplex (i.e. cannot transmit and
receive at the same frequency) single-antennas. Each data packet
$\Theta_i$ is error control coded and modulated, and transmitted
in $T$ time slots.

We assume the users transmit in equal time slots with transmit
power $\bar P$. The channel between any pair of nodes is assumed
to be frequency flat fading with additive white Gaussian noise
(AWGN). Let $ \textbf{x}_j \in \field{C}^{1 \times T}$ denote the
transmitted symbols from node $j$ and $\textbf{y}_i\in
\mathbb{C}^{1 \times T}$ the received symbols at node $i$, and the
additive noise has independent and identically distributed
(i.i.d.) entries $\textbf{z}_{i,t} \sim \mathbb{CN}(0,1)$, and
$\textbf{h}_{i,j} \in \field{C}$ the instantaneous channel
realization. Then, the channel within one block can be written as
\begin{equation}\label{systemmodel}
  \textbf{y}_i=\sqrt{\rho}\textbf{h}_{i,j}\textbf{x}_j+\textbf{z}_i
\end{equation}
where $\rho$ is the average received SNR at the destination. In
the above equation, the transmitter could be any of the sources or
relays, the receiver could be any of the relays or destinations,
as long as the transmitter and receiver are different (i.e., not
the same relay). We assume that the channel coefficient
$\textbf{h}_{i,j}$ remains constant during the transmission time
of a packet.

The channel coefficient $\textbf{h}_{i,j}$ between any two nodes
is modeled as i.i.d. with zero-mean, circularly symmetric complex
Gaussian random variables with common variance $1/\beta$.
Therefore, $|\textbf{h}_{i,j}|^2$ are exponentially distributed
with parameter $\beta$. A total of $NL$ bits are transmitted in
$(N+M)T$ channel uses, therefore the system rate is
$R=\frac{NL}{(N+M)T}$ bits per channel use (BPCU). The
transmission rate $R_i$ for one source or one relay per one packet
is fixed, identical, and equal to $R_i=\frac{L}{T}=\frac{N+M}{N}R$
BPCU.

\subsection{Deterministic Network Coded Cooperation (DNCC)} \label{main}

Our transmission scheme consists of two stages; see
\figref{fig.ta}(c). In the first stage, direct transmissions from
the sources to the destinations take place in $N$ orthogonal time
slots. Thanks to the broadcast nature of the wireless medium, all
the destinations and the relays overhear the transmissions. At the
end of the first stage, each relay tries to decode all $N$
packets. If a relay can successfully decode all the packets (we
denote this assumption by $\cal{A}$), then it participates in the
second stage. Otherwise, it remains silent.

In the second stage, the participating relays perform network
coding. Specifically, relay $i$ will transmit the linear
combination
\begin{equation}\label{linear comb}
    \sum_{k=1}^{N} \alpha_{ik} \Theta_k
\end{equation}
where $\Theta_k$ is the corresponding finite field element in
$\mathbb{F}_q$ for $\theta_k$.

The coefficients $\alpha_{ij}$'s are predetermined and they are
designed in a way to maximize the chance that the received linear
combinations are actually decodable at the destination. We discuss
the problem of how to choose these predetermined coefficients in
detail in \secref{design}.

In order to express the overall transmitted data packets, we
define the following matrix:
\begin{equation} \label{matr}
 A  := \begin{bmatrix}
     1 & 0 & . & . & 0 \\
     0 & 1 & . & . & 0 \\
     . & . & \ddots & . & . \\
     0 & . & . & . & 1 \\
     \alpha_{1,1} & \alpha_{1,2} & . & . & \alpha_{1,N} \\
     \alpha_{2,1} & . & . & . & \alpha_{2,N} \\
     . & . & . & . & . \\
     \alpha_{M,1} & \alpha_{M,2} & . & . & \alpha_{M,N}
 \end{bmatrix}
\end{equation}
where the top part is an $N\times N$ identity matrix. We also
define the $N\times1$ finite field vector corresponding to the
original source packets as
\begin{equation}
\Theta=[\Theta_1, \Theta_2, \ldots, \Theta_N]^T
\end{equation}
where $(\cdot)^T$ denotes transpose. Using matrices $A$ and
$\Theta$, the potential transmitted data packets can be expressed
as $X=A\Theta$.

\subsection{Design of the Linear Network Coding Matrix}
\label{design}

Each destination $d_i$ will attempt to decode all the packets that
have been transmitted at the two stages. It is possible that some
of the packets cannot be decoded due to severe channel fading.
However, if enough number of packets can be decoded, and the
linear network coding coefficients $\alpha_{m,n}$ are judiciously
designed, then the destination will be able to obtain the
transmitted packets by solving a linear system of equations. In
this subsection, we discuss the issue of the design of the coding
matrix $A$ (or the bottom $M$ rows of $A$ to be exact).

The row Kruskal-rank \cite{wagi01i} of $A$, denoted as
$\kappa(A)$, is the number $r$ such that every set of $r$ rows of
$A$ is linearly independent, but there exist one set of $r + 1$
rows that are linearly dependent. The column Kruskal-rank can be
defined similarly.


The following result relates the column Kruskal-rank of the
parity-check matrix of a linear block code to its minimum distance
$d_{min}$: The column Kruskal-rank $\kappa(H)$ of the parity check
matrix $H$ of a linear block error-control code and the minimum
Hamming distance $d_{min}$ of the code are related by
$d_{min}=\kappa(H) + 1$. Please refer to \cite{masl77} for a
proof.

We would like to have the matrix $A$ to have a large row Kruskal
rank so that a minimum number of equations is needed by any
destination to solve for the source packets. The row Kruskal rank
is of course less than or equal to the rank, which is $N$. To
maximize the Kruskal rank, we should design $A$ such that it has
Kruskal rank equal to $N$. Or stated differently, any $N$ rows of
the matrix $A$ should be linearly independent. Depending on the
sizes $N$ and $M$, this may or may not be possible in a given
finite field $\mathbb{F}_q$. Note that whether this is possible is
closely related whether maximum distance separable (MDS) codes
exists in a certain field for certain code dimensions
\cite{masl77}.

The transpose $H^T$ of the parity check matrix of a systematic
$(N+M,M,N+1)$ MDS code can be used as an encoding matrix $A$ for
our DNCC scheme to minimize the total number of packets necessary
at the destinations for decoding the source packets. If such an
$A$ is used, then each destination needs and only needs $N$
packets (from the sources and relays) for correct decoding.

\section{Diversity-Multiplexing Tradeoff} \label{sec.dmt}
\begin{figure}
\centering
\includegraphics[width=.30\textwidth]{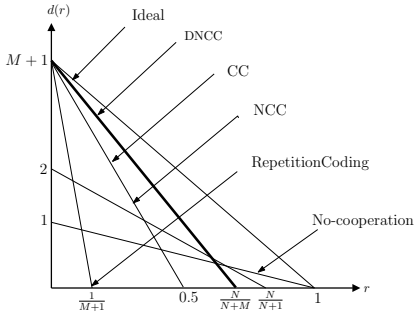}
\caption{DMT comparison of various schemes} \label{fig.dmt}
\end{figure}
As mentioned in the introduction, we will investigate the
performance of the proposed scheme via diversity-multiplexing
tradeoff (DMT). DMT is widely accepted as a useful performance
analysis tool in cooperative systems \cite{pzzy08,bkrl06}. For
completeness, we give the formal definitions as in \cite{zhts03}:
A scheme $C(\rho)$ is said to achieve spatial multiplexing gain
$r$ and diversity gain $d$ if the data rate is
\begin{equation}\label{rn}
  \lim_{\rho \rightarrow \infty} \frac{R(\rho)}{\log(\rho)} = r ,
\end{equation}
and the average error probability is
\begin{equation}\label{d}
  \lim_{\rho \rightarrow \infty} \frac{\log(P_e(\rho))}{\log(\rho)} = -d .
\end{equation}

\subsection{Outage Probability}

In \cite{zhts03}, for long enough block length $T$ it was shown
that the probability of error is dominated by the outage
probability. Therefore, we will only consider outage probabilities
in our analysis. The instantaneous mutual information of the
channel model in \eqref{systemmodel} is given by:
\begin{equation}\label{cap}
  I(\textbf{X}_i;\textbf{Y}_j)=\log(1+|\textbf{h}_{i,j}|^2\rho).
\end{equation}
where $\textbf{X}_i$ and $\textbf{Y}_j$ denote the transmitted
symbol by node $i$ and received symbol by node $j$. We write
$a(\tau)\cong b(\tau)$ if $\lim_{\tau\to 0} [a(\tau)/b(\tau)]=1$.
An outage event between a transmitter $i$ and receiver $j$ occurs
when the instantaneous mutual information for the channel
$h_{i,j}$ is less than the transmission rate $R_i$ BPCU. Since
$|\textbf{h}_{i,j}|^2$ is exponentially distributed, the outage
probability for the channel in \eqref{systemmodel} is given by:
 \begin{align}\label{out}
  P_0 & =Pr(I(\textbf{X}_i;\textbf{Y}_j)<R_i) \nonumber \\
      & =Pr(|\textbf{h}_{i,j}|^2<\tau)
      =1- \exp(-\beta\tau) \cong \beta\tau,
\end{align}
where $\tau = \frac{2^{\frac{N+M}{N}R}-1}{\rho}$.

\subsection{Further Improvements} \label{further}

\subsubsection{Decoding at the relays}
Decode-and-forward schemes suffer from performance loss when the
source-relay channel is in outage. And if a multi-source scenario
is considered the performance loss becomes even more severe.
Therefore, the assumption that the relay has to decode all the
packets in order to be able to cooperate becomes a bottleneck for
such schemes. We could relax the assumption $\cal{A}$ and assume
that the relays will participate cooperation even though they have
not been able to decode all the packets. Specifically, in the
second stage, the participating relays perform network coding on
the packets that they have received correctly. If relay $i$ was
able to decode the packets correctly from the sources in the set
$S_i$ where $S_i\subseteq \{1,\ldots, N\}$, then it will transmit
the linear combination
\begin{equation}\label{linear comb1}
    \sum_{k \epsilon S_i} \alpha_{i,k} \Theta_k
\end{equation}
where $\Theta_k$ is the corresponding finite field element in
$\mathbb{F}_q$ for $\theta_k$.

\begin{theorem} \label{th.maindmt}
The diversity-multiplexing tradeoff of the linear network coded
cooperation with $M$ intermediate relay nodes for both multicast
and unicast and both either with or without the assumption
$\cal{A}$ is:
\begin{equation}\label{dmtdncc}
 d\left(r\right) = \left(M+1\right)\left[1-\frac{\left(N+M\right)}{N}r\right],
  r \in \left(0,\frac{N}{N+M}\right)
\end{equation}
\end{theorem}
\begin{proof}
Due to severe fading some of the received packets may not be
successfully decoded by the destination. This can be viewed as
some of the rows of $A$ being erased. We denote the resulting
submatrix by $A_i$ for destination $d_i$. The successfully
received packets at destination $d_i$ after decoding can be
expressed as $Y_i= A_i\Theta$. In the multicast problem,
destination $d_i$ cannot recover $\Theta_i$ when the submatrix
$A_i$ is rank deficient, i.e. when $rank(A_i)<N$. This happens
when less than $N$ channels are not in outage which results in an
$A_i$ matrix that has at most $N-1$ rows. We define this event as
$E_{1i}=\{ A_i$ has at most $N-1$ rows\}.

Next, we define several other events which will simplify the
analysis. Let $E_m$ denote the event that $m$ relays fail to
receive all the $\Theta_i$'s correctly. Then we have:

\begin{eqnarray}\label{Ai}
 P(E_m) &=& \left( \begin{array}{c}
              M \\
              m
            \end{array} \right)  P(\varepsilon)^{M-m}
            (1-P(\varepsilon))^m  \nonumber \\
 P(\varepsilon) &=& \prod_{i=1}^{N} {Pr(I_{s_ir}(X;Y)>R_i) }
 \\
 &=& \prod_{i=1}^N Pr(|h_{s_ir}|^2 > \tau)
 = \exp(-N\beta\tau)
 \end{eqnarray}

Similarly, define $E(m,n)$ to be the event that $n$ channels out
of $m$ were in outage:
\begin{align}\label{E(m,n)}
 P(E(m,n)) = \left( \begin{array}{c}
              m \\
              n
            \end{array} \right)  P_0^{n}
            (1-P_0)^{m-n}
 \end{align}
where $P_0$ is given by \eqref{out}. Now, we can express
$P(E_{1i})$ as:
\begin{align}\label{error1i}
  P(E_{1i}) = \sum_{m=0}^{M} {P(E_m)} \cdot
  {\sum_{n=M-m+1}^{N+M-m} P(E(N+M-m,n))}
 \end{align}
The first summation stands for the probability of the event that
$m$ of the relays were in outage, leaving us with only $N+M-m$
nodes which were not in outage for that particular destination.
Notice that, as $\rho \rightarrow \infty$, $\tau \rightarrow 0$.
We would like to find the following limit:
  \begin{equation}\label{limit}
 \lim_{\tau \rightarrow 0}{\frac{P(E_{1i})}{\tau^{M+1}}}
  \end{equation}

We consider the individual terms in the summations one-by-one and
find the term with the smallest order of $\tau$. Observe that
$P(E_m) \cong K_m\tau^m $ where $K_m$ is a constant:
 \begin{eqnarray}\label{limitEm}
 && \lim_{\tau \rightarrow 0}{\frac{P(E_m)}{\tau^{m}}}
=      \left( \begin{array}{c}
              M \\
              m
             \end{array} \right)
             \exp(-N(M-m)\beta\tau)
             \nonumber \\
&& \cdot \lim_{\tau \rightarrow 0} \left\{   \frac{
(1-\exp(-N\beta\tau))^m}{\tau^{m}}\right\}
              = \left( \begin{array}{c} M \\ m \end{array} \right)(N\beta)^m
  \end{eqnarray}

Next, consider the term $P(E(m,n))$,
  \begin{eqnarray}\label{limitsum}
   \lim_{\tau \rightarrow 0} {\frac{P(E(m,n))}{\tau^{n}}} &=&
  \lim_{\tau \rightarrow 0}{\frac{{\left( \begin{array}{c}
              m \\
              n
            \end{array} \right) {P_0^{n} (1-P_0)^{m-n}}}}
{\tau^{n}} } \nonumber \\
 & = &\left( \begin{array}{c}
              m \\
              n
            \end{array} \right)\beta^{n}
  \end{eqnarray}

When $n$ is equal to $M-m+1$, we have the smallest order $\tau$
term . Substituting $n=M-m+1$ in $P(E(N+M-m,n))$, we have:
  \begin{eqnarray*}\label{limitsum1}
  && \lim_{\tau \rightarrow 0} \{
  \left( \begin{array}{c}
              N+M-m \\
              M-m+1
            \end{array} \right)  (1-P_0)^{N+M-m-(M-m+1)}
            \nonumber \\
     &&         \cdot           \frac{P_0^{M-m+1}}
{\tau^{M-m+1}}  \}
  = \left( \begin{array}{c}
              N+M-m \\
              M-m+1
            \end{array} \right)\beta^{M-m+1}
  \end{eqnarray*}

Therefore, we have:
  \begin{align}\label{limitoverall}
  & \lim_{\tau \rightarrow 0}
  \frac{P(E_m)}{\tau^{m}}\cdot
   {\frac{P(E(N+M-m,n))}{\tau^{M-m+1}}}
\nonumber \\
 & = \binom Mm (N\beta)^m
            \binom{N+M-m}{M-m+1}
            \beta^{M-m+1}
  \end{align}
Finally, using \eqref{limitoverall}, we have:
  \begin{align}\label{overall}
&P(E_{1i})  \cong  \tau^{M+1} \sum_{m=0}^{M} \binom Mm N^m
\binom{N+M-m}{M-m+1}
            \beta^{M+1} \nonumber \\  &=
\left({\frac{2^{\frac{N+M}{N}R-1}}{\rho}}\right)^{M+1}
                      \sum_{m=0}^{M} \binom Mm \binom{N+M-m}{M-m+1} N^m\beta^{M+1}
   \end{align}
Now, suppose the fixed rate has chosen to be:
   \begin{align} \label{coderate}
   R=r\log \rho
   \end{align}
Substituting \eqref{coderate} into \eqref{overall}, we arrive at:
   \begin{eqnarray}\label{final}
 P(E_{1i}) & \cong & \rho^{(\frac{N+M}{N}r-1)(M+1)}
\nonumber \\ & \cdot & \sum_{m=0}^{M} \binom Mm
            \binom{N+M-m}{M-m+1}
            N^m \beta^{M+1  }
   \end{eqnarray}
Since the summation term is just a constant in terms of $\rho$, we
have the desired result. We skip the proof for the unicast and
without assumption $\cal{A}$ cases due to space limitations.
\end{proof}

\section{Comparison with Other Schemes} \label{sec.compare}
\subsection{DMT Comparison}
\begin{figure*}
\subfigure[\vspace{-.1in}  ]{\label{fig.casea}
\includegraphics[width=.34\textwidth]{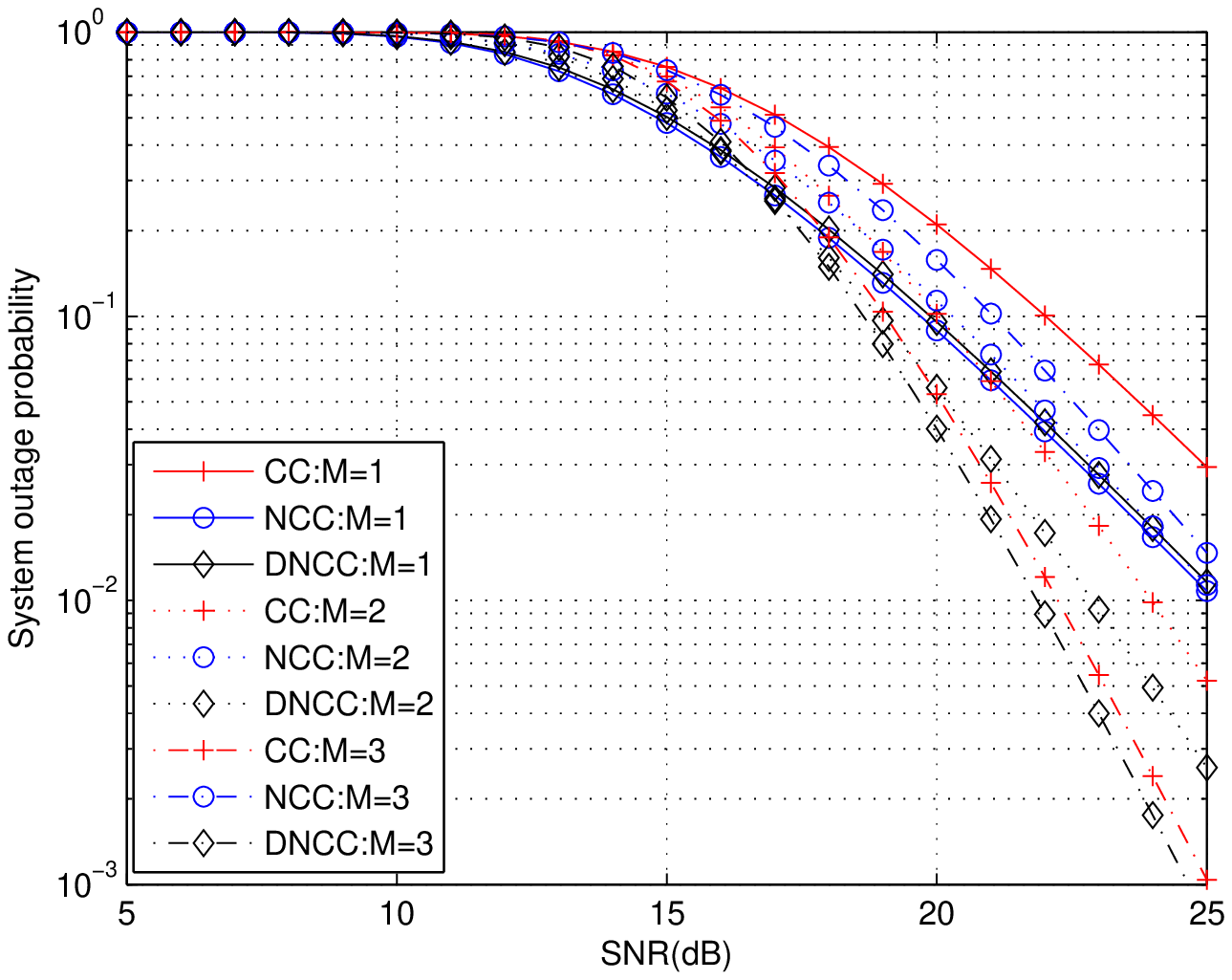}}
\subfigure[\vspace{-.1in} ]{\label{fig.caseb}
\includegraphics[width=.34\textwidth]{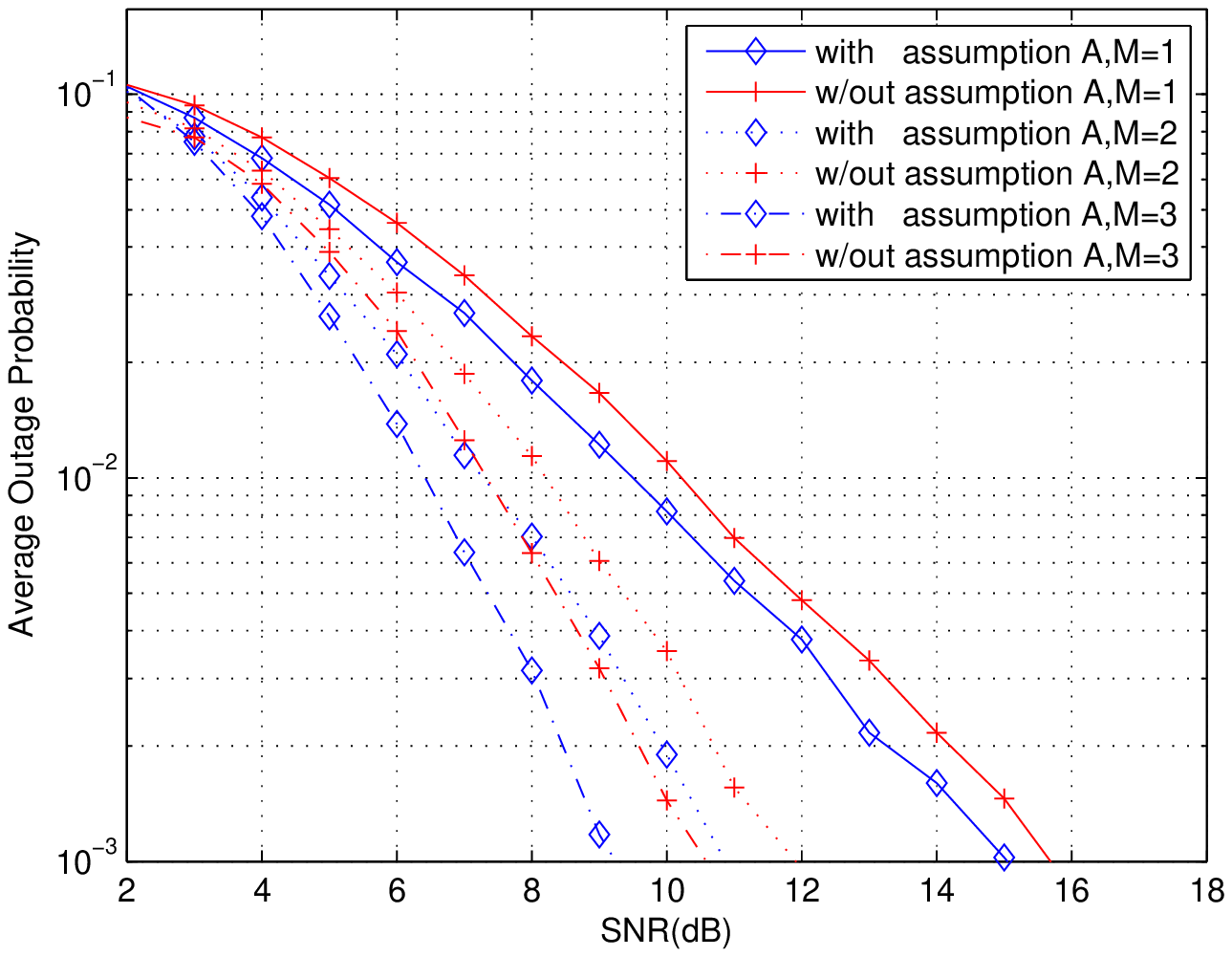}}
\subfigure[\vspace{-.1in}  ]{\label{fig.casec}
\includegraphics[width=.34\textwidth]{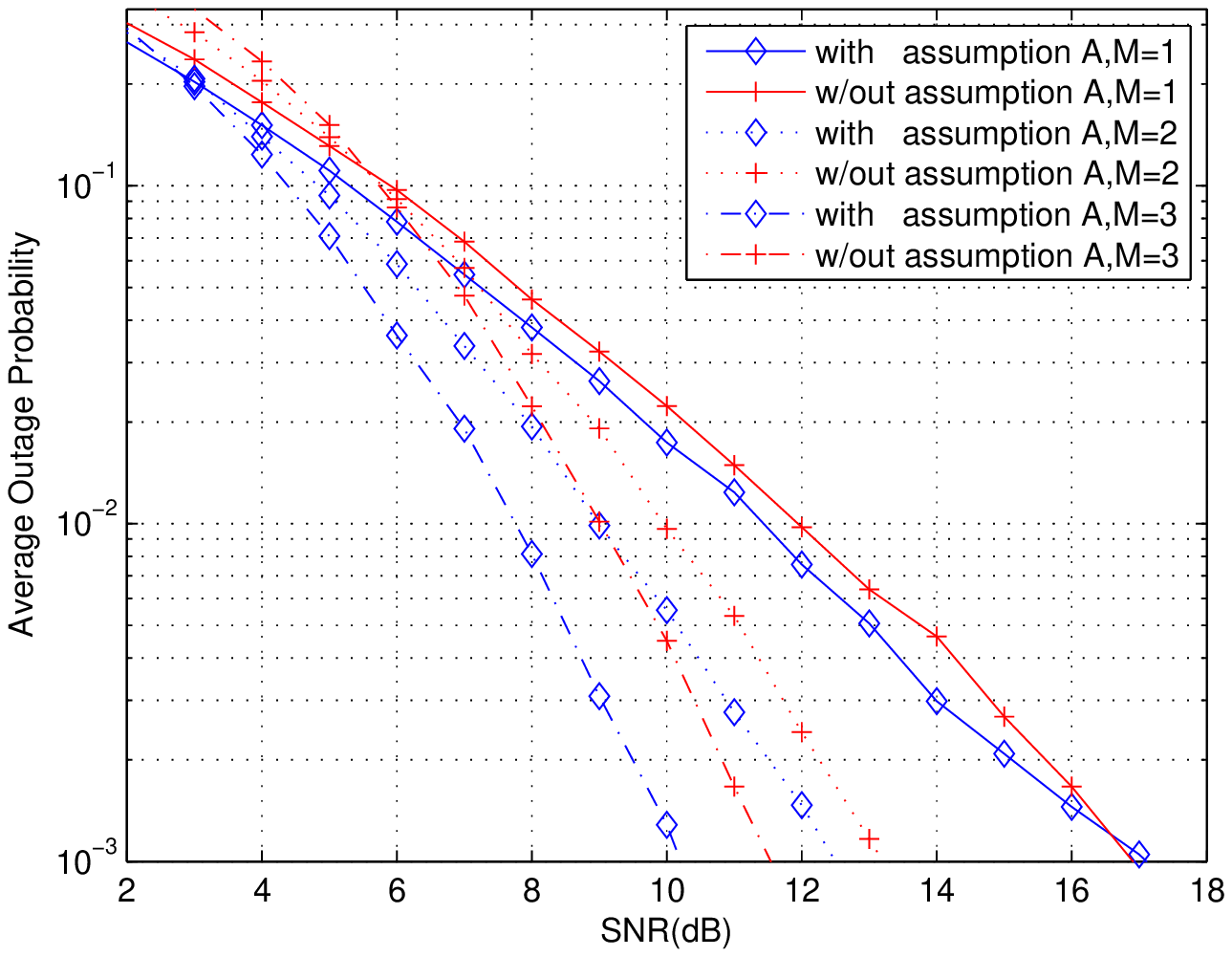}}
\caption{System outage probability (N=2 M=1, \ldots 3), Average
outage probability per destination, (N=2 M=1, \ldots 3 and N=3
M=1, \ldots 3) } \label{fig.plots}
\end{figure*}

In this section, we would like to compare the proposed scheme with
the existing schemes in the literature in terms of
diversity-multiplexing tradeoff. The closest scheme in the
literature is the Network-coded cooperation (NCC) considered in
\cite{pzzy08}. There are $N$ s-d pairs and $M$ relays in a single
cell. However, instead of all the relays, only one relay transmits
following the direct transmissions from the source nodes, which
results in total of $N+1$ time slots (\figref{fig.ta} (b)). Using
fewer time-slots NCC achieves a better spectral efficiency than
DNCC. However, NCC can only provide a fixed diversity order of 2,
while the proposed scheme achieves the full-diversity order of
$M+1$.

In the following, for comparison we include the DMT performance of
NCC and that of conventional cooperation (CC).

The diversity-multiplexing tradeoff of NCC is given by
\cite{pzzy08}:
\begin{equation}\label{NCC}
d\left(r\right) = 2\left(1-\frac{N+1}{N}r\right) , \quad r \in
\left(0,\frac{N}{N+1}\right)
\end{equation}

The DMT of the CC schemes with $M$ intermediate relay nodes is
given by \cite{bkrl06,lawo03}:
\begin{equation}\label{CC}
d\left(r\right) = \left(M+1\right)\left(1-2r\right) , \quad r \in
\left(0,0.5\right).
\end{equation}

We present diversity-multiplexing tradeoff of the existing schemes
and DNCC in \figref{fig.dmt}. As can be seen from the figure,
although DNCC can achieve the full diversity order, NCC fails to
do so. We also see that DNCC achieves a better DMT than CC when $N
> M$.

\subsection{System Outage Probability}

Here, we compare the system outage probability of DNCC with the
other schemes. The system outage occurs when \emph{any} $d_i$ is
unable to decode $\Theta_i$ reliably:

\begin{equation}\label{sysout}
P_s = 1-\prod_{i=1}^N{(1-P(E_{1i}))}
\end{equation}

We compare \eqref{sysout} with the system outage probabilities
(30), (43) derived in \cite{pzzy08}. In \figref{fig.casea} the proposed
method clearly outperforms NCC by achieving the full diversity of
$M+1$ as compared to NCC's fixed diversity order of 2.

\subsection{Monte-Carlo Simulation}

We also compare the proposed scheme with the existing schemes via
Monte-Carlo simulations. In the simulations, only channel
conditions are considered to isolate the diversity benefits of the
scheme. We generate an $(N+M) \times N$ and an $N \times N$ matrix
that contains the channel coefficients for each destination and
each relay, respectively. Then, we decide that the transmission is
successful for any link if the instantaneous channel condition is
large enough to be able to support the given data rate and we
update the same size linear coefficient matrices accordingly.
After all the transmissions take place, we perform Gaussian
elimination on the updated linear coefficient matrices to conclude
whether each destination $d_i$ was able to recover the source
packet $\Theta_i$ or not. The channel coefficient variances are
chosen to be equal to one and $R_i=1$ BPCU. Please note that we
considered the average outage error probability which is found by
dividing the total number of errors occurred by the number of
source nodes instead of the system error probability. In all the
figures only the unicast scenario is adapted since CC cannot be
implemented in a multicast scenario. As it is proven by the
\thref{th.maindmt}, the performance loss incurred due to the
assumption $\cal{A}$ is not in terms of diversity gain but it is
in terms of coding gain. This is validated through simulations as
shown in the \figref{fig.caseb} and \figref{fig.casec}.

\section{Conclusions} \label{sec.conclusion}

In this paper, we have proposed a network coded cooperation scheme
for $N$ source-destination pairs assisted with $M$ relays. We
studied two different transmission scenarios, unicast and
multicast. The proposed scheme allows the relays to apply network
coding on the data it has received from its neighbors. We
establish the link between the parity-check matrix for a
$(N+M,M,N+1)$ MDS code and the coefficients to perform network
coding in a cooperative communication scenario consisting of $N$
source-destination pairs and $M$ relays. We also obtained the
diversity-multiplexing tradeoff performance of the proposed
scheme, and showed its advantage over the existing schemes.
Specifically, it achieves the full-diversity order $M+1$ at the
expense of a slightly reduced multiplexing rate.

\bibliographystyle{IEEE}
\bibliography{refs}

\end{document}